\documentclass[twocolumn,tighten,times]{aastex62}

\usepackage{amssymb,amsmath}
\usepackage{graphicx}
\usepackage{color}
\usepackage{subfigure}
\usepackage{lineno}

\newcommand{\caii}{\ion{Ca}{2}}

\newcommand{\mgii}{\ion{Mg}{2}}
\newcommand{\fei}{\ion{Fe}{1}}
\newcommand{\feia}{\ion{Fe}{1}~6302~\AA}
\newcommand{\feib}{\ion{Fe}{1}~6173~\AA}

\newcommand{\iris}{{\em IRIS}}

\newcommand{\dkist}{{\em DKIST}}
\newcommand{\visp}{{\em ViSP}}

\received{}
\submitjournal{ApJ}

\shorttitle{}
\shortauthors{Mart\'inez-Sykora et al.}

\newcommand{\longacknowledgment}{We gratefully acknowledge support by NASA grants 80NSSC20K1272, 80NSSC21K0737, 80NSSC21K1684, NSF grant AST1714955 and contract NNG09FA40C (\iris). Resources supporting this work were provided by the NASA High-End Computing (HEC) Program through the NASA Advanced Supercomputing (NAS) Division at Ames Research Center. The simulation has been run on Pleiades cluster through the computing project s1061, s2601. This research is also supported by the Research Council of Norway through its Centres of Excellence scheme, project number 262622, and through grants of computing time from the Programme for Supercomputing. Data are courtesy of \iris. \iris\ is a NASA small explorer mission developed and operated by LMSAL with mission operations executed at NASA Ames Research Center and major contributions to downlink communications funded by ESA and the Norwegian Space Centre.}

\begin{document}

\title{Small-scale Chromospheric Spectropolarimetric Observables in the Internetwork from a 3D Radiative MHD Model}

\correspondingauthor{Juan Mart\'inez-Sykora}
\email{martinezsykora@baeri.org}

\author[0000-0002-0333-5717]{Juan Mart\'inez-Sykora}
\affil{Lockheed Martin Solar \& Astrophysics Laboratory, 3251 Hanover Street, Palo Alto, CA 94304, USA}
\affil{Bay Area Environmental Research Institute, NASA Research Park, Moffett Field, CA 94035, USA}
\affil{Rosseland Centre for Solar Physics, University of Oslo, P.O. Box 1029 Blindern, NO-0315 Oslo, Norway}
\affil{Institute of Theoretical Astrophysics, University of Oslo, P.O. Box 1029 Blindern, NO-0315 Oslo, Norway}

\author[0000-0002-3234-3070]{Alberto Sainz Dalda}
\affil{Lockheed Martin Solar \& Astrophysics Laboratory, 3251 Hanover Street, Palo Alto, CA 94304, USA}
\affil{Bay Area Environmental Research Institute, NASA Research Park, Moffett Field, CA 94035, USA}

\author[0000-0002-5879-4371]{Milan Go\v{s}i\'{c}}
\affil{Lockheed Martin Solar \& Astrophysics Laboratory, 3251 Hanover Street, Palo Alto, CA 94304, USA}
\affil{Bay Area Environmental Research Institute, NASA Research Park, Moffett Field, CA 94035, USA}

\author[0000-0002-8370-952X]{Bart De Pontieu}
\affil{Lockheed Martin Solar \& Astrophysics Laboratory, 3251 Hanover Street, Palo Alto, CA 94304, USA}
\affil{Rosseland Centre for Solar Physics, University of Oslo, P.O. Box 1029 Blindern, NO-0315 Oslo, Norway}
\affil{Institute of Theoretical Astrophysics, University of Oslo, P.O. Box 1029 Blindern, NO-0315 Oslo, Norway}

\begin{abstract}

The presence of the magnetic field is critical to transport energy through the solar atmosphere. The new generation of telescopes will provide new insight into how the magnetic field arrives into the chromosphere and its role in the energy balance of the solar atmosphere. We have used a 3D radiative MHD numerical model of the solar atmosphere with high spatial resolution ($\sim4$~km) calculated with the Bifrost code. This code solves the full MHD equations with non-grey and non-LTE radiative transfer and thermal conduction along magnetic field lines. The model shows how the lower chromosphere in the internetwork, a region dominated by magneto-acoustic shocks and where plasma beta is greater than 1, is able to generate magnetic field in-situ  \citep{Martinez-Sykora:2019dyn}. We have synthesized full-polarimetric Stokes profiles from this model for several spectral lines formed in the photosphere and the chromosphere. These synthetic profiles illustrate the types of observables expected from \dkist\ and \iris . Our work provides insight into how to interpret observations from these observatories. We find that in order to discern the chromospheric magnetic observables it is crucial to compensate for Doppler shift rather than use fix wavelength range. 
 
\end{abstract}


\section{Introduction}

The solar surface outside of active regions, known as the quiet Sun (QS), is permeated by magnetic fields concentrated along supergranular boundaries, called the network (NE), and small and weak internetwork (IN) magnetic fields in the interior of supergranular cells. Despite their small polarization signals, IN is established as a highly dynamic environment in which magnetic features are observed to continually appear, disappear, and accumulate at the boundaries of supergranular cells where they form network fields \citep{Gosic:2014ef, Gosic:2016vk, Gosic:2022ApJ...925..188G}. This renders IN features as an essential contributor to the solar magnetism since they completely shape the QS magnetic carpet on granular and supergranular scales. Due to their abundance, IN fields store a huge amount of magnetic flux and energy in the solar photosphere, which could in principle, if those fields reach the chromosphere,  balance the radiative losses at chromospheric heights \citep{Trujillo:2004Natur.430..326T}.

Magnetic fields generated in the photosphere may act as the main agent to channel Poynting flux to the upper solar atmosphere. However, it is unknown how these fields are formed, to what extent they are capable of expanding into the chromosphere and beyond, and how IN fields vary with height. From the observational perspective, there are indications that the footpoints of the rising loops produce localized brightenings in the chromosphere \citep{Martinez-Gonzalez:2010kb, Kontogiannis:2020AA...633A..67K}. In addition to this, \cite{Gosic:2021ApJ...911...41G} provided direct observational evidence that at least the strongest IN fields can be observed at chromospheric and transition region heights. However, the general contribution and presence of IN magnetic fields in the chromosphere is still unknown. It is clear that further studies are required to obtain a solution to the debate about the presence and nature of IN fields in the photosphere, including their impact on the solar atmosphere. High signal-to-noise observations are required to properly determine the magnetic field vector of these structures which show such a weak polarization signal. In addition, high spatial resolution is key to minimize the Zeeman cancellation of the circular polarization signals. Such observations can be obtained with the Daniel K. Inouye Solar Telescope \citep[DKIST;][]{Elmore2014SPIE.9147E..07E}.

At the surface of the Sun, the local dynamo time-scales are shorter and the magnetic energy growth larger than in deeper layers of the convection zone \citep[e.g.,][]{Nordlund:2008dq,Vogler:2007yg,Rempel:2014sf,Kitiashvili:2015nr,Khomenko:2017sf}. As mentioned above, current observations seem to indicate that the small-scale photospheric field rarely penetrates the chromosphere. The very low rate of small-scale emergence into the chromosphere may agree with model predictions of a strong negative Poynting flux at the surface \citep[e.g.,][]{Abbett:2012kc} compatible with the magnetic field returning to below the surface through the intergranular lanes. Other suggested  mechanisms may play a role in the Poynting flux at the surface. The Hall term could lead to changes in the vertical Poynting flux, but those changes seem small \citep{Gonzalez-Morales:2020AA...642A.220G,Khomenko:2021RSPTA.37900176K}. Integrating over the simulated box, the Hall term may play a role in the propagation of Alfv\'en waves rather than introducing a new magnetic field in the chromosphere. A recent numerical model by \citet{Martinez-Sykora:2019dyn} suggests that chromospheric QS or CH internetwork dynamics may be able to transform kinetic energy into magnetic energy. In this work, we analyze how this chromospheric magnetic field may be observed, for instance, with \dkist. We also show how our synthetic observables could be used to interpret the observations. Ultimately, of course, the \iris\ \citep{De-Pontieu:2014yu} and \dkist\ observations will be critical to constrain our models.

In the following, the numerical model is briefly described in Section~\ref{sec:mod}. Section~\ref{sec:syn} explains the synthesis and the methods to analysis of the Stokes profiles, and Section~\ref{sec:obs} the selected \iris\ observations. The results are shown in Section~\ref{sec:res}. One subsection focuses on the synthetic Stokes profiles and magnetic field diagnostics (Section~\ref{sec:sto}), while the second subsection describes the intensities and provides a comparison with \iris\ and FTS Atlas observations  (Section~\ref{sec:res_obs}). Finally, the manuscript ends with a conclusion and discussion (Section~\ref{sec:dis}). 

\section{Numerical model}~\label{sec:mod}

In this effort, we use the 3D numerical model described in detail in \citet{Martinez-Sykora:2019dyn}. In short, this 3D radiative-MHD numerical simulation is computed with the Bifrost code \citep{Gudiksen:2011qy}. The model includes radiative transfer with scattering in the photosphere and lower chromosphere \citep{Skartlien2000,Hayek:2010ac}. In the middle and upper chromosphere, radiation from specific species such as hydrogen, calcium, and magnesium is computed following the recipes of \citet{Carlsson:2012uq}. We use optically thin radiative losses in the transition region and corona. We also include thermal conduction along the magnetic field, which is important for the energetics of the transition region and corona. 

The simulation spans a vertical domain stretching from $\sim2.5$~Mm below the photosphere to $8$~Mm above into the corona with a non-uniform vertical grid size of $4$~km in the photosphere and chromosphere. The photosphere is located at $z=0$, where $\tau_{500} \sim 1$. The horizontal domain spans $6\times6$~Mm in the $x$ and $y$ directions with $5$~km resolution. Initially, the simulation box is seeded with a uniform weak vertical magnetic field of $2.5$~G. Then, the convective motions introduce and build up magnetic field complexity, and the magnetic field strength ultimately reaches a statistical steady state with $B = 56$~G at photospheric heights \citep[similar to that described by][]{Vogler:2007yg,Rempel:2014sf,Cameron:2015rm}. 

Figure~\ref{fig:vercut} shows a vertical cut of the magneto-thermo-dynamic structure at $t=4250$~s. The magnetic field is accumulated in the downflows in the convection zone ($z<0$). Shocks dominate the simulated chromosphere ($0.5<z<2$~Mm) and jets ($2<z<4$~Mm), whereas the corona is maintained with heating from magnetic braiding and dominated by thermal conduction. The analysis/synthesis is done for the snapshot at $t=4250$~s. By this instance, the simulation has reached a steady state. 

In this high-resolution QS model, the chromospheric magnetic energy content increases substantially with time, especially in the region where shocks dominate ($0.5<z<2$~Mm). This magnetic energy growth is local to the chromosphere and mostly caused by the plasma dynamics in this high plasma beta region. Note that the chromospheric magnetic growth is due to colliding shocks and strong velocity shear, instead of the turbulent motion of the convective cells in the convection zone. Further details of this model can be found in \citet{Martinez-Sykora:2019dyn}.

\section{Synthesis of Stokes profiles}~\label{sec:syn}

We compute the synthetic Stokes profiles using the RH radiative transfer code \citep{Uitenbroek:2001dq,Pereira:2015th} considering polarization due to the Zeeman effect and the field-free approximation. This approach starts with iterations of the non-LTE solution with unpolarized profiles, followed by a formal solution with polarization, but only after convergence. The synthesis of the \caii~H and the \mgii~k lines considers partial frequency redistribution (PRD). The optically thick synthesis was calculated individually in every tenth vertical column ($\mu=1$) of the simulation domain. Note that this sampling is close to \dkist\ spatial resolution \citep[0.03-0.05\arcsec][]{Rimmele2020SoPh..295..172R}. We compute the synthetic Stokes profiles for \feia, \feib, \mgii\ UV, \caii~8542 \AA, \caii~H, and \mgii~k. 

A useful diagnostic is to compute the mean circular polarization degree (MCPD) and mean linear polarization degree (MLPD), which give us valuable information about the magnetic field where the Stokes profiles were generated. The MLPD is calculated as in \citet{Sainz-Dalda:2012qf}:

\begin{equation}
MLPD = \left(\int_{\lambda_0}^{\lambda_1}{\frac{\sqrt{Q^2(\lambda)+U^2(\lambda)}}{I(\lambda)}}d\lambda \right)/(\lambda_1-\lambda_0)
\end{equation}

while the MCPD is:

\begin{equation}
MCPD = \left(\int_{\lambda_0}^{\lambda_1}{\frac{|V(\lambda)|}{I(\lambda)}}d\lambda \right)/(\lambda_1-\lambda_0)
\end{equation}

Another common diagnostic tool is the weak-field approximation, which is valid when the Zeeman splitting of a line ($g_{eff}\Delta\lambda_B$, with $g_{eff}$ the effective Landé) is much smaller than its Doppler broadening ($\Delta\lambda_D$) Thus, the longitudinal component of the magnetic field can be derived from the following expression:

\begin{equation}
V(\lambda) = - \Delta \lambda_B g_{eff} \cos \theta \frac{dI(\lambda)}{d\lambda}
\end{equation}

with $\theta$ being the inclination of the magnetic field, $V(\lambda)$ and $I(\lambda)$ the Stokes V and Stokes I profiles respectively, and

\begin{equation}
\Delta \lambda_B = 4.6686 \times 10^{-13} B \lambda_0^2
\end{equation}

with $\lambda_0$ the center of the spectral line in rest given in \AA. 

\section{\iris\ Observations}~\label{sec:obs}
\begin{figure}[!t]
    \centering
    \resizebox{1\hsize}{!}{\includegraphics[width=0.88\textwidth]{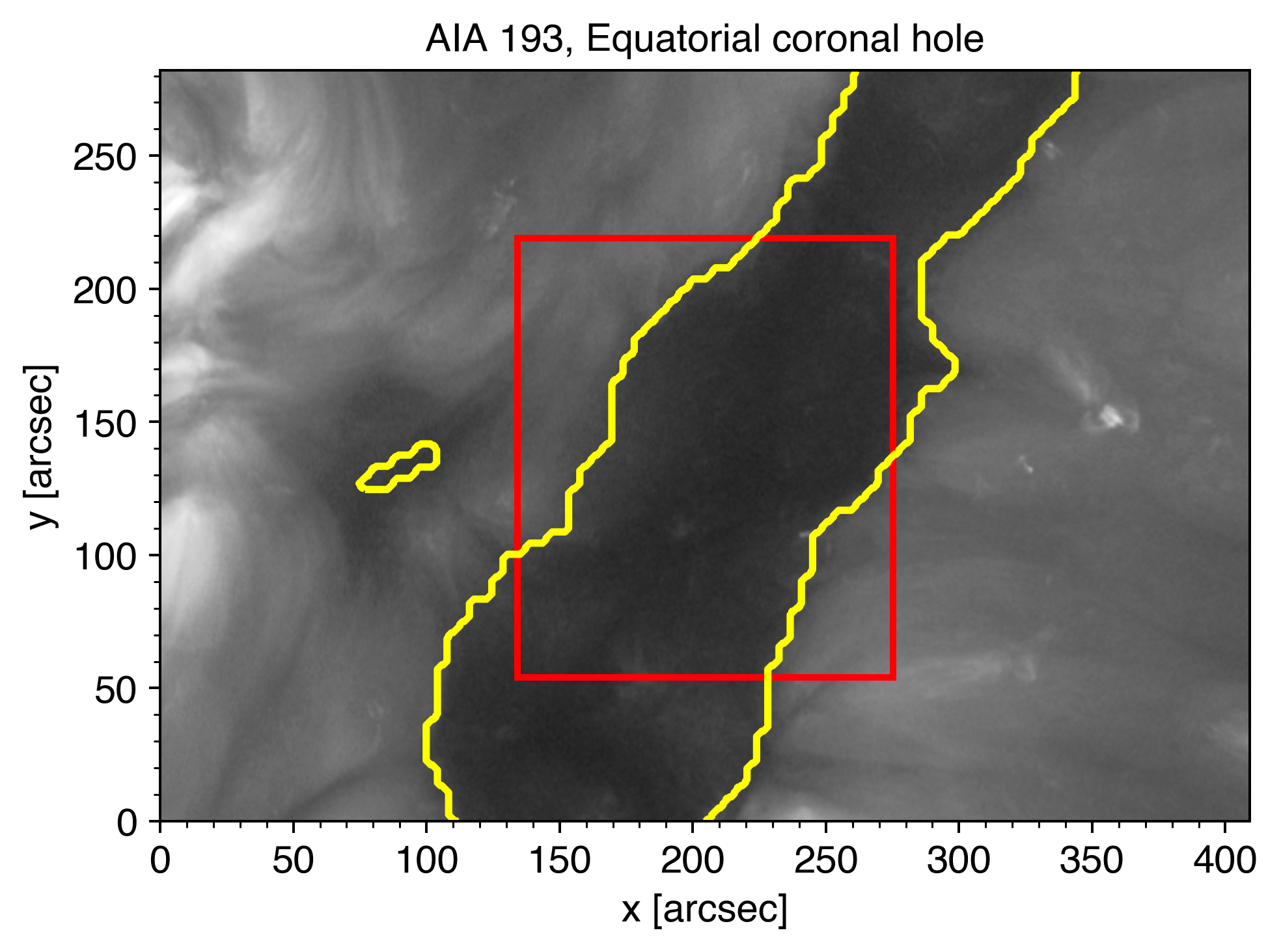}}
    \caption{AIA filtergram at 193 Å showing the selected equatorial coronal hole (yellow contour). The red box outlines the area covered by the IRIS slit. The coronal hole was identified using the CHRONNOS method \citep{2021A&A...652A..13J}}
    \label{fig:ch_iris_fov_aia}
\end{figure}

The observations used in this work were obtained by \iris\ \citep{De-Pontieu:2014yu} on 2016 March 25, and 2017 October 15. The former (data set 1) started at 10:09:18~UT and ended at 11:58:45~UT. The latter (data set 2) observed the Sun from 17:54:22~UT until 22:52:35~UT. They were both taken at  disk center. Data set 1 shows the evolution of an equatorial CH as shown with the 193 AIA image in Figure~\ref{fig:ch_iris_fov_aia}, while data set 2 monitored a QS region.

Data set 1 is a very large dense 400-step raster (0.33\arcsec\ step size), taking spectra of several spectral lines in two far ultraviolet (FUV) domains that range from $1332$~\AA\ to $1358$~\AA\ and from $1391$~\AA\ to $1407$~\AA\/. In the near ultraviolet (NUV) band, \iris\ obtained spectra within a subset of the wavelength range that covers 2790~\AA\ to 2835~\AA\/. Here we consider only the NUV \ion{Mg}{2} h \& k spectral lines that sample the solar atmosphere from the photosphere up to the upper chromospheric layers. The cadence of the spectral observations is 16~s per raster step (the total raster cadence is 110 minutes), covering a CH region of $130\arcsec\times175$\arcsec. Slit-jaw images were taken using the \ion{C}{2} 1330~\AA\ (SJI 1330), \ion{Si}{4} 1400~\AA\ (SJI 1400), \ion{Mg}{2} k 2796~\AA\ (SJI 2796), and \ion{Mg}{2} h wing at 2832~\AA\ (SJI 2832) filters, without solar rotation tracking. In all cases the pixel size is $0\farcs16$. The cadence of the slit-jaw images at 1300~\AA\/, 1400~\AA\/, and 2796~\AA\/ is 68~s. The slit-jaw images at 2832~\AA\ are recorded at a cadence of 410~s. 

Data set 2 is a large dense 96-step raster (0.33\arcsec\ step size) over a QS region. The spectra are taken every 60 seconds, with the total duration of one raster repeat about 100 minutes. The FOV of this raster is $30\arcsec\times120$\arcsec, and accounts for solar rotation. The slit-jaw filters are the same as for data set 1, with a pixel size of 0.16\arcsec. The cadences of the slit-jaw images are 199~s for SJI 1330, 1400, and 2796. The SJI 2832 images have a cadence of about 995~s. 

The \iris\ observations were corrected for dark current, flat-field, geometric distortion, and scattered light \citep{De-Pontieu:2014yu,Wulser2018SoPh..293..149W}. The calibration pipeline also includes wavelength calibrations and subtraction of the background light leak in FUV data. Finally, we performed radiometric calibration to convert the Data Number units of the measured intensities into W~m$^{-2}$~sr$^{-1}$~Hz$^{-1}$.

\section{Results}~\label{sec:res}

First, we will characterize the magnetic field properties in the lower atmosphere (photosphere and chromospheric) in locations where our chosen spectral lines are formed. Integrated synthetic stokes profiles and MLPD and MCPD observables will be compared with the magnetic field at the location of the line formation of the selected spectral lines. We will calculate the weak field approximation (WFA) within narrow spectral windows around several lines to provide better magnetic field observables and compare those with the magnetic field in the model. In the subsequent section, we will compare the model predictions with the observed chromospheric intensity profiles, highlighting the model's strengths and limitations. 

\subsection{Stokes profiles, WFA and magnetic observables}\label{sec:sto}

\begin{figure*}
    \centering
    \includegraphics[width=0.9\textwidth]{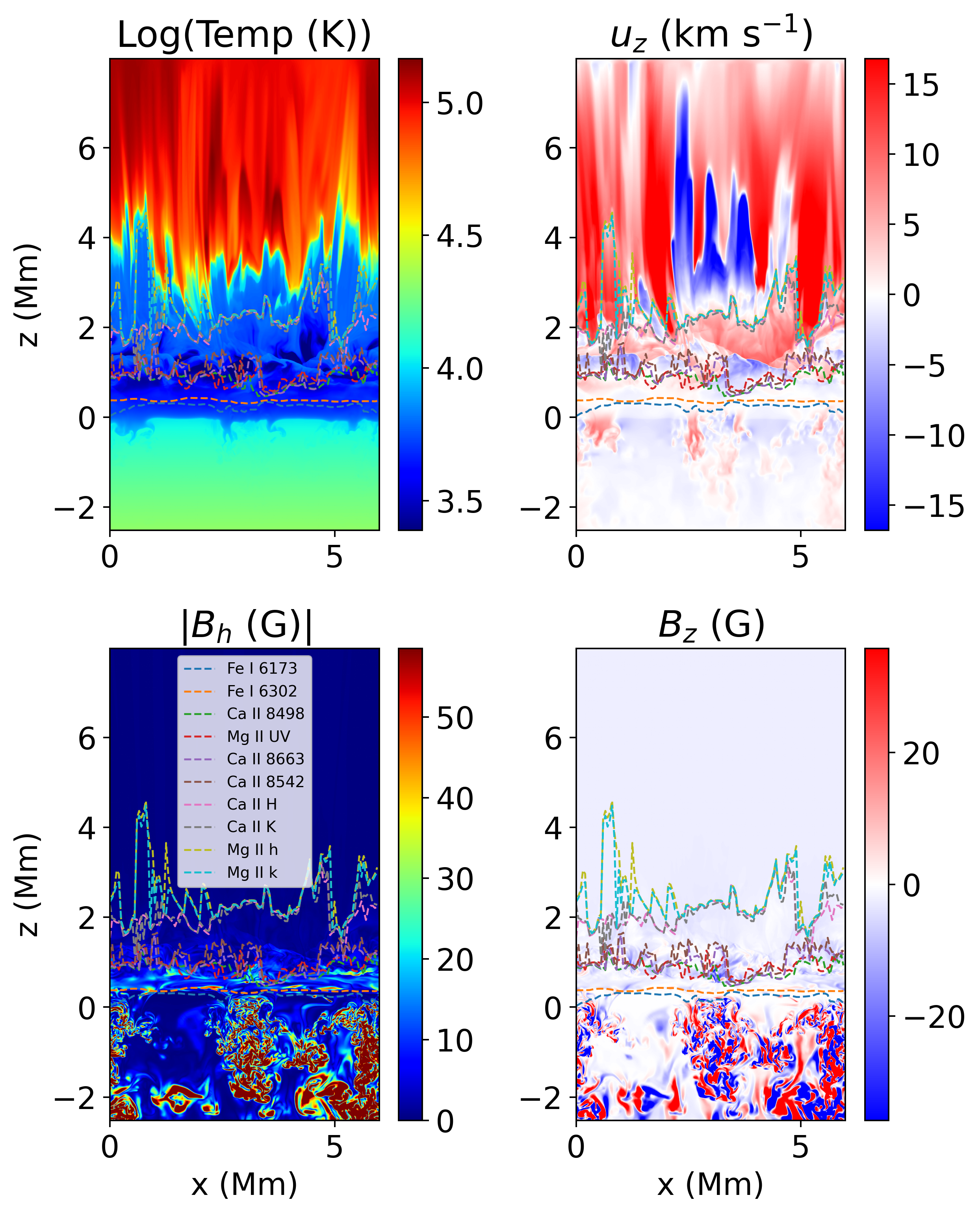}
    \caption{From top to bottom: Maps of a vertical cut showing temperature, vertical velocity, and horizontal and vertical field strength. We show with dashed lines in the bottom left panel the height for an optical depth ($\tau$) equal to unity at the core of the lines listed in the legend of the bottom-left panel. In the convection zone ($z<0$), the magnetic field is accumulated in the downflows. Shocks dominate the simulated chromosphere ($0.5<z<2$~Mm) and jets ($2<z<4$~Mm). The corona is maintained by heating from magnetic braiding and dominated by thermal conduction.}
    \label{fig:vercut}
\end{figure*}

The formation height where the selected lines (\feia, \feib, \mgii~UV, \caii~IR triplet, \caii~H/K, and \mgii~h/k) are sensitive to the atmosphere's physical conditions covers a corrugated region that covers a wide range of heights in the solar atmosphere. It is usual to calculate the height where the optical depth equals one for the line of interest to indicate where the formation region is as a reference.
Figure~\ref{fig:vercut} includes the heights for an optical depth ($\tau$) equal to unity at the core of the chosen lines in the numerical domain. Both \feia\ and \feib\ lines are located in the upper photosphere, i.e., in the region where granules overshoot and where the photospheric field is squeezed and very frequently moved to the downflows and returned into the convection zone \citep{Martinez-Sykora:2019dyn}. Going up through the atmosphere, the following spectral lines are \mgii~UV and \caii~IR triplet (middle chromosphere). At these heights the plasma is dominated by gas pressure ($\beta > 1$) with acoustic shocks traveling in all directions. This is also the region where the chromospheric magnetic local growth (with time) occurs. In the upper chromosphere, we find the \caii~H, K, \mgii~h, and k lines. These lines are above the cold chromospheric bubbles and are mostly located in spicules. In comparison with \citet{Leenaarts:2013hc}, \caii~IR triplet seems to be formed at lower heights in our model, whereas \caii~K and \mgii~k are formed at greater heights. The former difference is most likely due to the fact that \citet{Leenaarts:2013hc} included non-equilibrium hydrogen ionization. The latter is due to longer spicules in this model compared to \citet{Leenaarts:2013hc}. Note that the \citet{Leenaarts:2013hc} model has a much larger spatial pixel size (i.e., lower resolution), and the magnetic field configuration is very different and representative of QS with two opposite (network-like) flux concentrations separated by 8~Mm. In addition, that model included the effects of non-equilibrium ionization of hydrogen \citet{Leenaarts:2012cr}. 

Another aspect to highlight is that the height of $\tau=1$ occurs over a height range of tens or a few hundred kilometers for photospheric lines and up to a few mega-meters for chromospheric spectral lines. Moreover, this height range increases with the formation height, e.g., the region where the $\tau=1$ height of \caii\ lines is smaller than for \mgii\ lines.

\begin{figure*}
    \centering
    \includegraphics[width=0.98\textwidth]{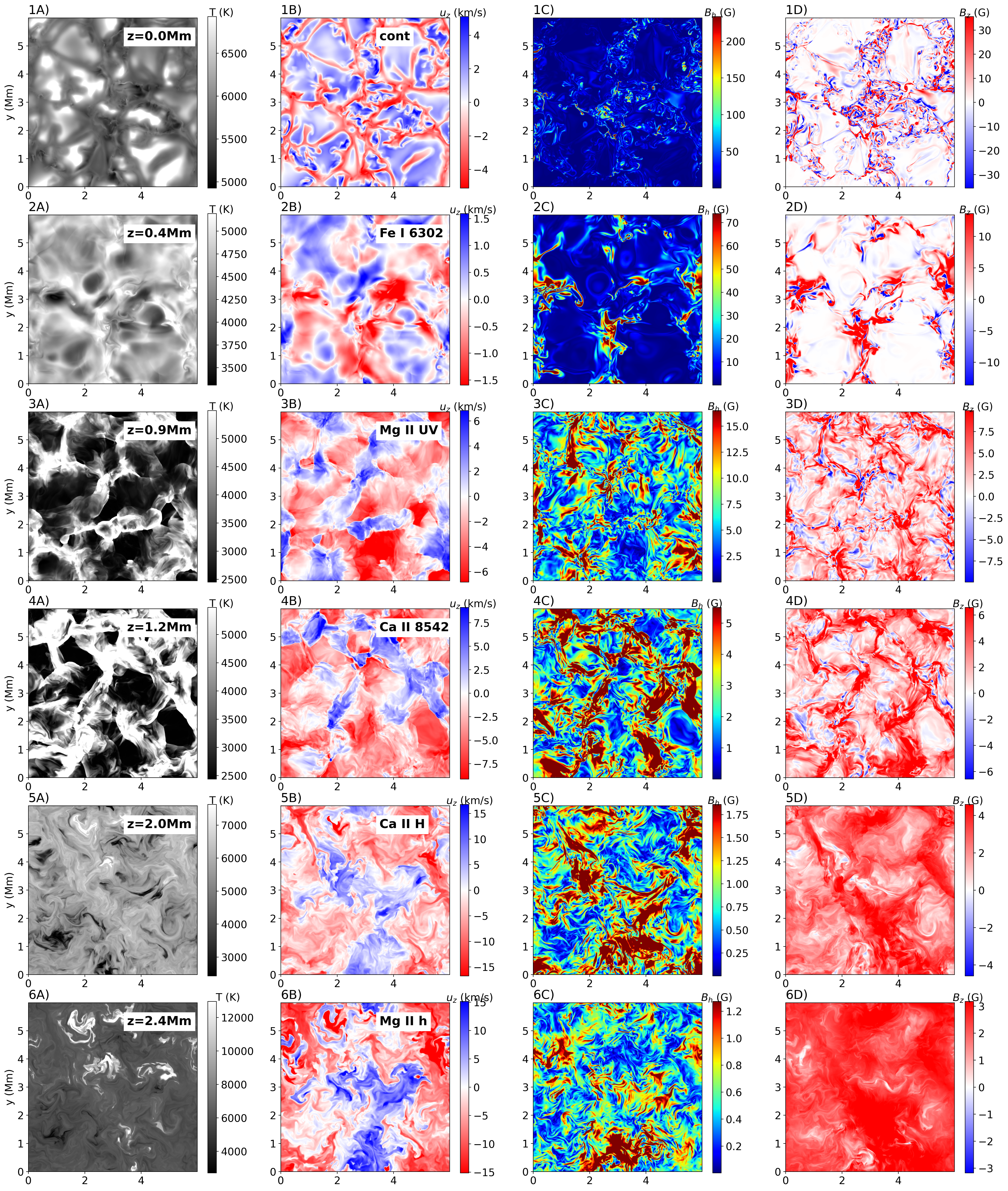}
    \caption{From left to right: maps of horizontal cuts of temperature, vertical velocity, and horizontal and vertical field strength at the average formation height formation of each line (shown at the top-right corner of the plots in the first column), i.e., from top to bottom continuum, \feia, \mgii~UV, \caii~8498, \caii~8542~\AA, \caii~H, and \mgii~h. Contrary to the photosphere, the flows and shocks are not that well correlated with the magnetic field configuration.}
    \label{fig:horcut}
\end{figure*}

Magneto-thermodynamic maps of horizontal cuts at the average $\tau=1$ heights of the various lines reveal that the dynamics and magnetic field configuration strongly correlate in the photosphere (Figure~\ref{fig:horcut}). At the continuum heights, the temperature, velocity, and magnetic field show the usual photospheric thermal properties in QS, i.e., ``wide" upflow granules with weaker fields and greater temperatures than the narrow intergranular lanes. In other words, the magnetic field is concentrated in the down flow regions, i.e., in the intragranular lanes. The high resolution reveals the substructure within the granules and intergranular lanes. Note the salt-and-pepper pattern on very small scales of the vertical magnetic field.  Temperature and velocity maps at the averaged $\tau=1$ height of \feib\ reveal the reverse granulation, and the magnetic field is distributed over wider regions along the line-of-sight (LOS). In addition, the magnetic field has lost most of the salt-and-pepper pattern. The magnetic field is mainly concentrated in the downflows, but not always, e.g., $(x,y)=(4,6)$~Mm. In the lower-mid chromosphere, i.e., at the averaged $\tau=1$ height for \mgii~UV, \caii~IR, the temperature and velocity maps show the thermodynamic properties of the acoustic shocks traveling (and colliding) in all directions. The magnetic field has lost most of the connectivity with the photosphere and one can no longer recognize the granulation pattern. In addition, the magnetic field is not concentrated in the downflows as in the photosphere, but it tends to be located in the region with gradients in velocity. Finally, it is very interesting that the magnetic field reveals smaller structures and more salt-and-pepper pattern than in the reverse granulation region. This pattern results from the dynamic properties of the chromosphere, which can produce a local growth of the magnetic field due to folding and reconnecting field lines. Finally, in the upper chromosphere, i.e., at the averaged $\tau=1$ height for \caii~H and \mgii~h, the temperature maps show some incursions from the transition region, i.e., high temperatures and the structure is more dominated by magnetic pressure and spicules forming in the mid-upper chromosphere. The velocity has no sharp discontinuities. The magnetic field is highly vertical, i.e., dominated by the initial constant value, with small perturbations on the horizontal field, and the salt-and-pepper pattern is gone.

\begin{figure}
    \centering
    \includegraphics[width=0.49\textwidth]{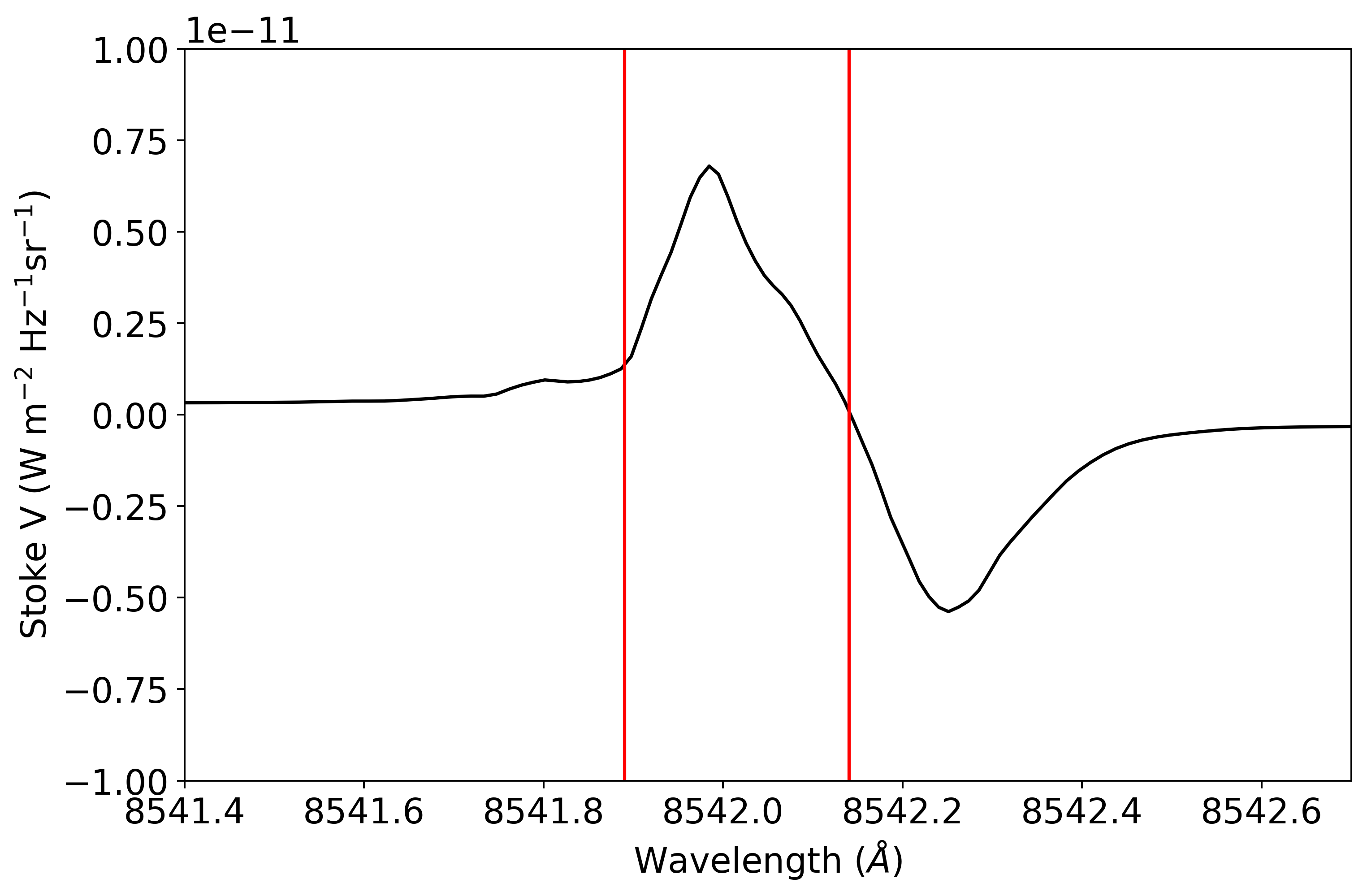}
    \caption{We consider the blue lobe for the stokes maps and WFA calculations. This example is for the mean Stoke V profile of the whole simulation of \caii~8542~\AA. The integration width is between the two red vertical lines and is 0.25~\AA.}
    \label{fig:meanvprof}
\end{figure}

\begin{figure*}
    \centering
    \includegraphics[width=0.99\textwidth]{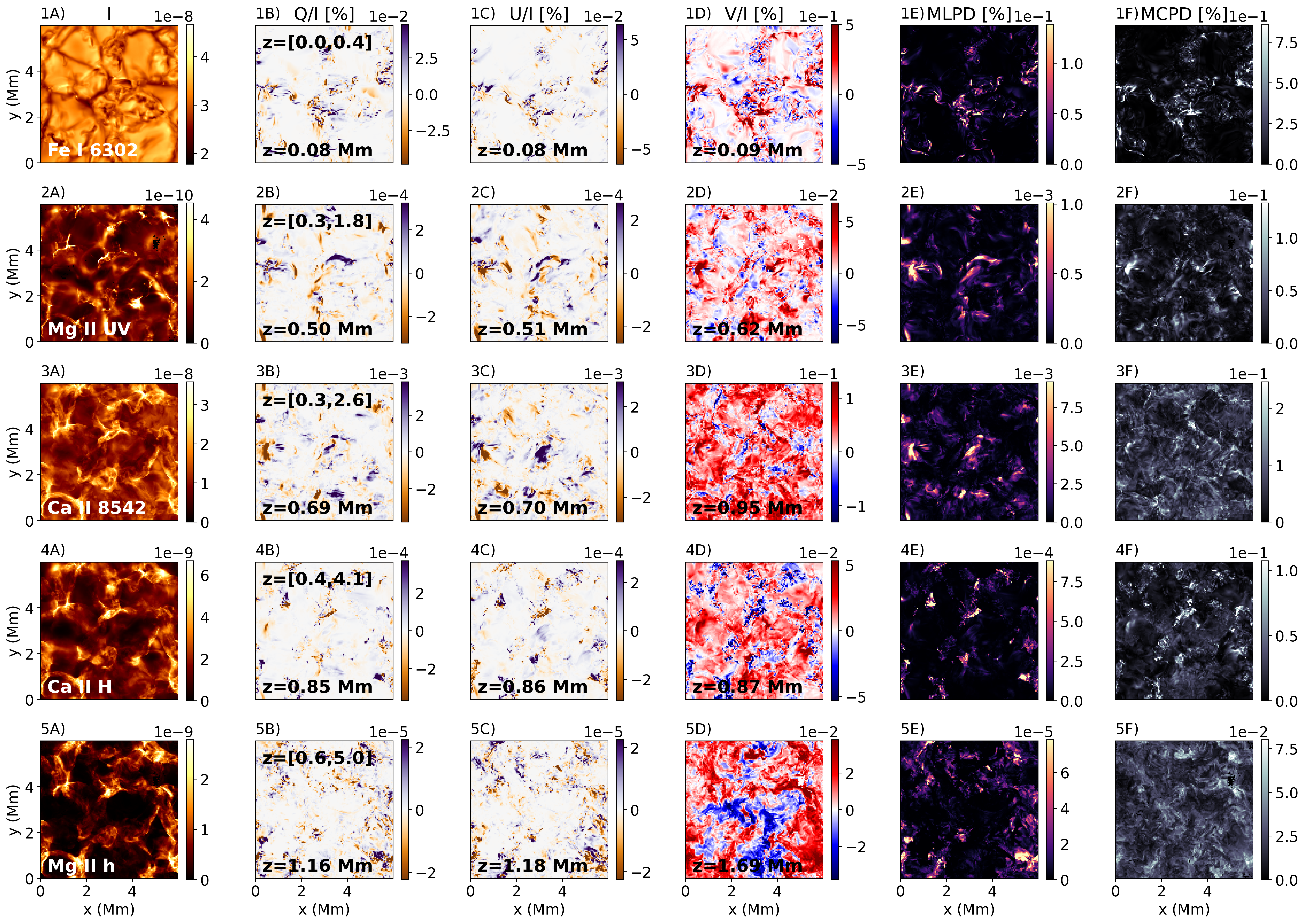}
    \caption{From left to right, integrated Stokes profiles (I, Q, U and V), mean linear polarization degree (MLPD), and mean circular polarization degree (MCPD) are shown, from top to bottom, for \feia, \mgii~UV, \caii~8542~\AA, \caii~H, and \mgii~k. The integrated intensity is given in W~m$^{-2}$Hz$^{-1}$sr$^{-1}$. The integrated Stokes, MLPD, and MCPD are given in \%. See the main text for further details.}
    \label{fig:mpnd}
\end{figure*}

Synthetic observables and the mean linear and circular polarization degree (MLPD, MCPD), for \feia, \feib, \mgii~UV, \caii~8542 \AA, \caii~H, and \mgii~k are shown in Figure~\ref{fig:mpnd}. These synthetic Stokes diagnostics could help to interpret upcoming \dkist\ observations. The first column shows the integrated intensity in the near continuum for the photospheric lines (\feia, and \mgii~UV) and the integrated intensity within the wavelength range around the line core wavelengths (as deduced from the average profile) for chromospheric lines (\caii~8542 \AA, \caii~H, and \mgii~k). The second, third, and fourth columns show the integrated Stokes maps in the wavelength range of the blue lobe of the mean Stokes $S(\lambda)$ profile; this is with $S(\lambda)$ being $Q(\lambda)$, $U(\lambda)$ and $V(\lambda)$ as shown with the example in Figure~\ref{fig:meanvprof}. The wavelength ranges are 0.22, 0.16, 0.25, 0.187, 0.134~\AA\ for \feia, \mgii~UV, \caii~8542~\AA, \caii~H, and \mgii~k, respectively. The fifth and fourth columns are the MLPD, and the MCPD calculated as explained in Section 3. For each panel in the second column of Figure~\ref{fig:mpnd} we include (inset towards the top)  the height range of $\tau=1$ over the integrated wavelength region, and at the bottom, its mean height is weighted with the corresponding Stokes profiles of $\tau=1$ as follows: 

\begin{eqnarray}
H=\frac{\int_{\lambda_o}^{\lambda_1} \tau(\lambda) S(\lambda) d\lambda}{\int_{\lambda_o}^{\lambda_1}S(\lambda) d\lambda}~\label{eq:height}
\end{eqnarray}

While for \fei, the range is a few hundred km, and the mean is roughly at 80~km above the photosphere, for \mgii~UV the height range is $\sim$ 1,000 km and its median is 500~km above the photosphere, i.e., the lower chromosphere. This is also similar for \caii\ IR. Finally, \caii\ H and \mgii~h/k are 1~Mm above the photosphere and occur over a height range of several Mm. 

The Stokes maps of \fei\ Q, U, and V, and MCPD, and MLPD reveal the photospheric magnetic field, as one can notice by comparing panels 1C, 1D, 2C, and 2D in Figure~\ref{fig:horcut} with the top row in Figure~\ref{fig:mpnd}. However, for the integrated Stokes maps formed at greater heights, it is not clear that we can resolve the magnetic field (compare panels 3D, 4D, 5D and 6D of Figure~\ref{fig:horcut} with panels 2-5 and B-F in Figure~\ref{fig:mpnd}). For example, with a close look at Stokes V, one can find a striking match between the vertical velocities (e.g., panels 5B and 6B in Figure~\ref{fig:horcut}) with the Stokes V map in panel 5D in Figure~\ref{fig:mpnd}. This is because we have performed the integration over a fixed wavelength range, as commonly done by observers, with the wavelength range around the core taken from the mean profiles. Because of the large Doppler shifts present in the simulation, this washes out the magnetic sensitivity of the Stokes observables.  It is also interesting to see that the Stokes V in the photosphere matches very well with the magnetic field and in the chromosphere with the vertical velocity.

\begin{figure*}
    \centering
    \includegraphics[width=0.99\textwidth]{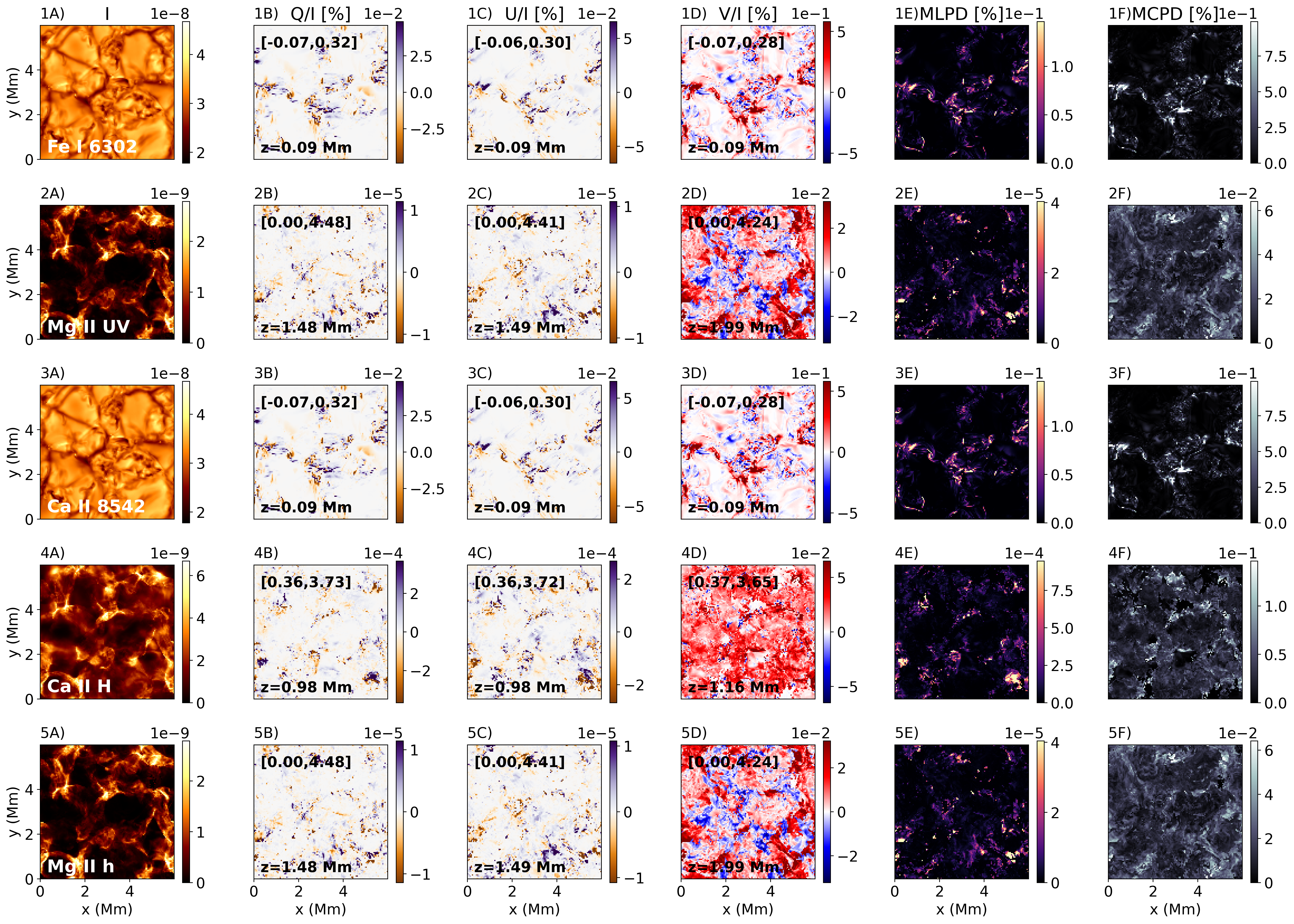}
    \caption{Equivalent to Figure~\ref{fig:mpnd} but the integration of the line core has been performed pixel by pixel. See the main text for further details.}
    \label{fig:mpd}
\end{figure*}

To better capture the chromospheric magnetic observables, Figure~\ref{fig:mpd} is equivalent to the previous one. However, instead of integrating over a fixed wavelength range within the core of the mean profiles, the line core wavelength is selected separately for each pixel within the FOV. This reduces the Doppler shift effects. The impact is minor in photospheric lines but large in chromospheric lines. Note the darkening of regions in the lower chromosphere (\caii\ and \mgii) due to shocks passing through. It is interesting to see that Q, U, and V Stokes, MCPD, and MLPD intensities dominate at the intergranular lanes in the photosphere. The chromosphere seems to have magnetic signals in regions close to bright areas in intensity. Those differences between the photosphere and chromosphere result from the lack of connection between the magnetic field in the photosphere to that of the chromosphere. Instead, in our simulation the chromosphere can maintain the magnetic energy and field from local conversion of kinetic energy into magnetic energy. Note that the signal is large enough and \dkist\ will have the capability to provide enough resolution and signal to constrain the results from this simulation and provide the chromospheric magnetic field observables, at least for Stokes V. The comparison of the Stokes and MCPD and MLPD maps with the magnetic field cuts in the right column of Figure~\ref{fig:horcut} shows some similarities, but still has large differences. One of the main reasons is that we are still integrating over a wavelength range and the Stokes maps are not real horizontal cuts as shown in Figure~\ref{fig:vercut} but follow the corrugated surface of $\tau=1$. 

\begin{figure}
    \centering
    \includegraphics[width=0.49\textwidth]{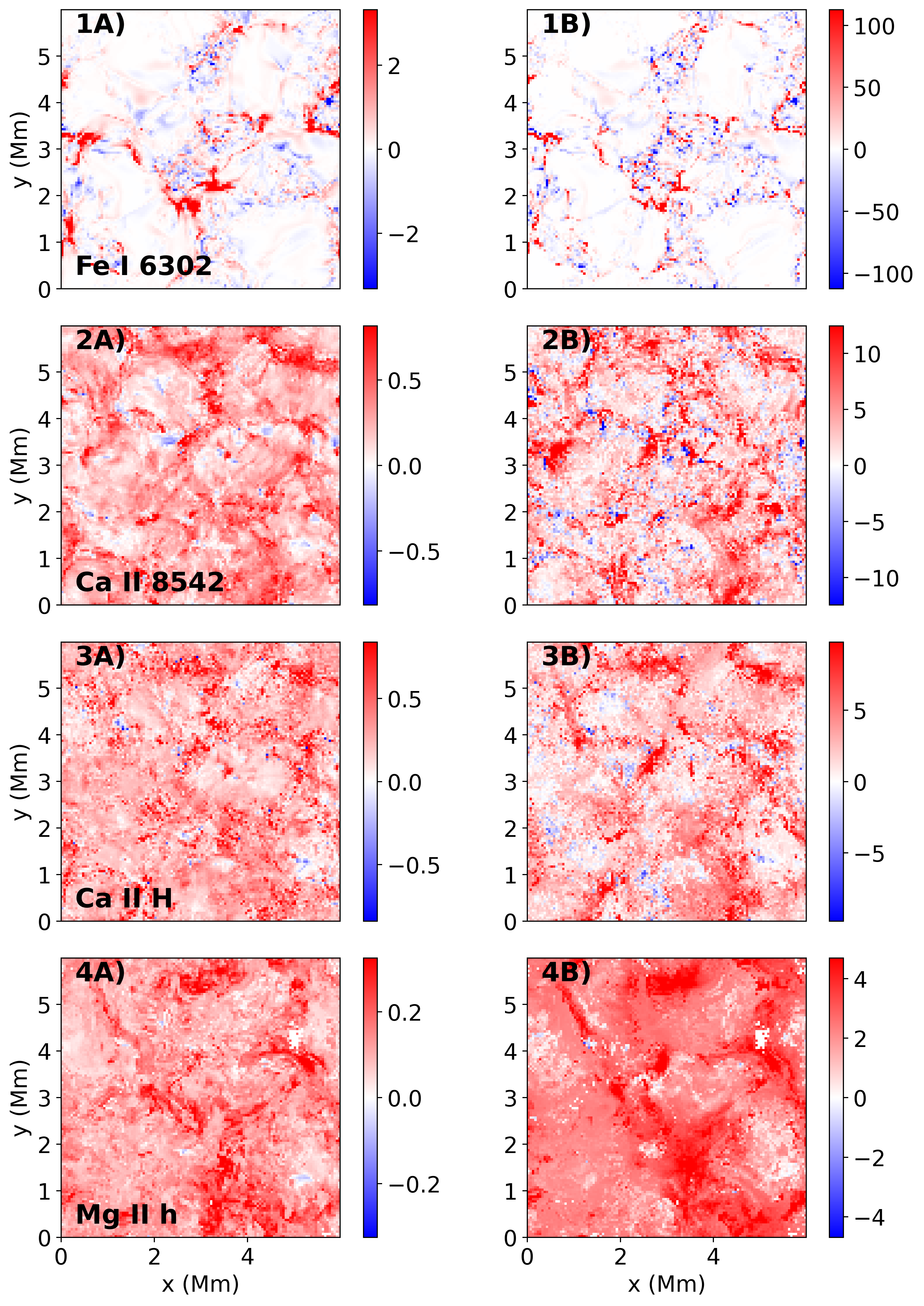}
    \caption{The LOS component of the magnetic field applying the WFA at the core of the spectral lines \fei, \caii~8542\AA, \caii~H, and \mgii\ are shown in the left column from top to bottom, respectively (left column). The vertical component of the magnetic field at the mean height for each pixel of $\tau=1$ within the core of the lines weighted with the intensity profiles are shown in the right column.}
    \label{fig:wfa}
\end{figure}

What then is needed to obtain a good estimate of the magnetic field in the chromosphere from these types of observables? We show that the WFA method is significantly better than the above measures.
Figure~\ref{fig:wfa} shows a comparison between the LOS component of the magnetic field ($B_{LOS}$) - applying the WFA to the core of the synthetic profiles - (left column) and the vertical component of the magnetic field of the numerical model ($B_{LOS}$) at the mean height of $\tau=1$ within the selected wavelength range of the line core for each FOV pixel weighted with intensity (right column) following Equation~\ref{eq:height} pixel by pixel in the FOV. Because the synthetic profiles were computed considering a view from above ($\mu=1$), $B_{LOS}$ is equal to $B_{z}$. For the chromospheric lines, the WFA was applied in the line core, avoiding the photospheric wings for each FOV taking into account Doppler shifts. The latter, if they had been included, would introduce photospheric information in the magnetic field  \cite[e.g.,][]{Centeno:2018ApJ...866...89C,AfonsoDelgado2020_prep}.

It is not surprising that the $B_{LOS}$ recovered from the WFA applied to the photospheric lines (e.g., \feia\ top row) matches well with $B_{z}$ in the expected formation heights of the line. The WFA signal is more diffuse because it is integrated over a range of heights. Whereas, the magnetic field is interpolated from the mean height $\tau=1$ for the right column of the figure. For the chromosphere, the WFA values reproduce rather well the magnetic features at those heights because we have appropriately selected a narrow region within the spectral line core as suggested by \citet{AfonsoDelgado2020_prep}. Consequently, these synthetic and magnetic field maps can be used to interpret \dkist\ observations, constrain this model, and identify if these physical processes, i.e., local chromospheric magnetic growth, occur in the solar atmosphere and, if so, for which targets on the Sun. We notice remarkable similarities between the magnetic field maps in the chromosphere (panels 4-6D in Figure~\ref{fig:horcut}, the magnetic field in panels 2-3B in Figure~\ref{fig:wfa}, and the WFA maps in panels 2-3A in Figure~\ref{fig:wfa}. 

\subsection{Comparison of synthetic chromospheric observables with observations}\label{sec:res_obs}

\begin{figure*}
    \centering
    \includegraphics[width=0.45\textwidth]{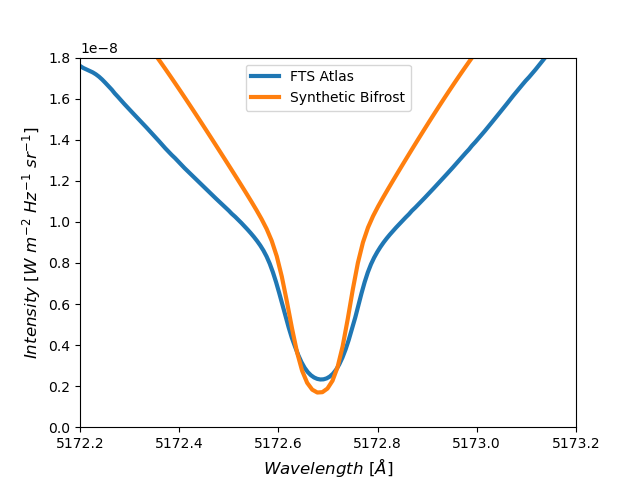}
    \includegraphics[width=0.45\textwidth]{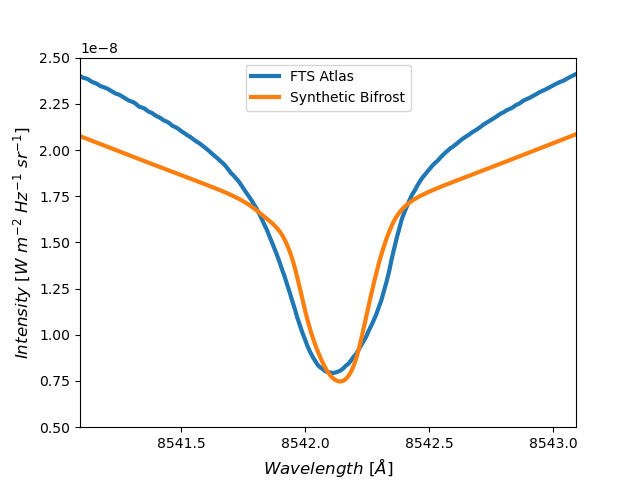}
    \caption{Comparison of FTS Atlas observations (blue) with the synthetic (orange) \caii~8542 (left), and H (right) averaged profiles reveals a relatively good match at the core of the lines, still a bit narrower for the synthetic profiles.}
    \label{fig:fts}
\end{figure*}

\begin{figure*}[!t]
    \centering
    \resizebox{1\hsize}{!}{\includegraphics[width=0.95\textwidth]{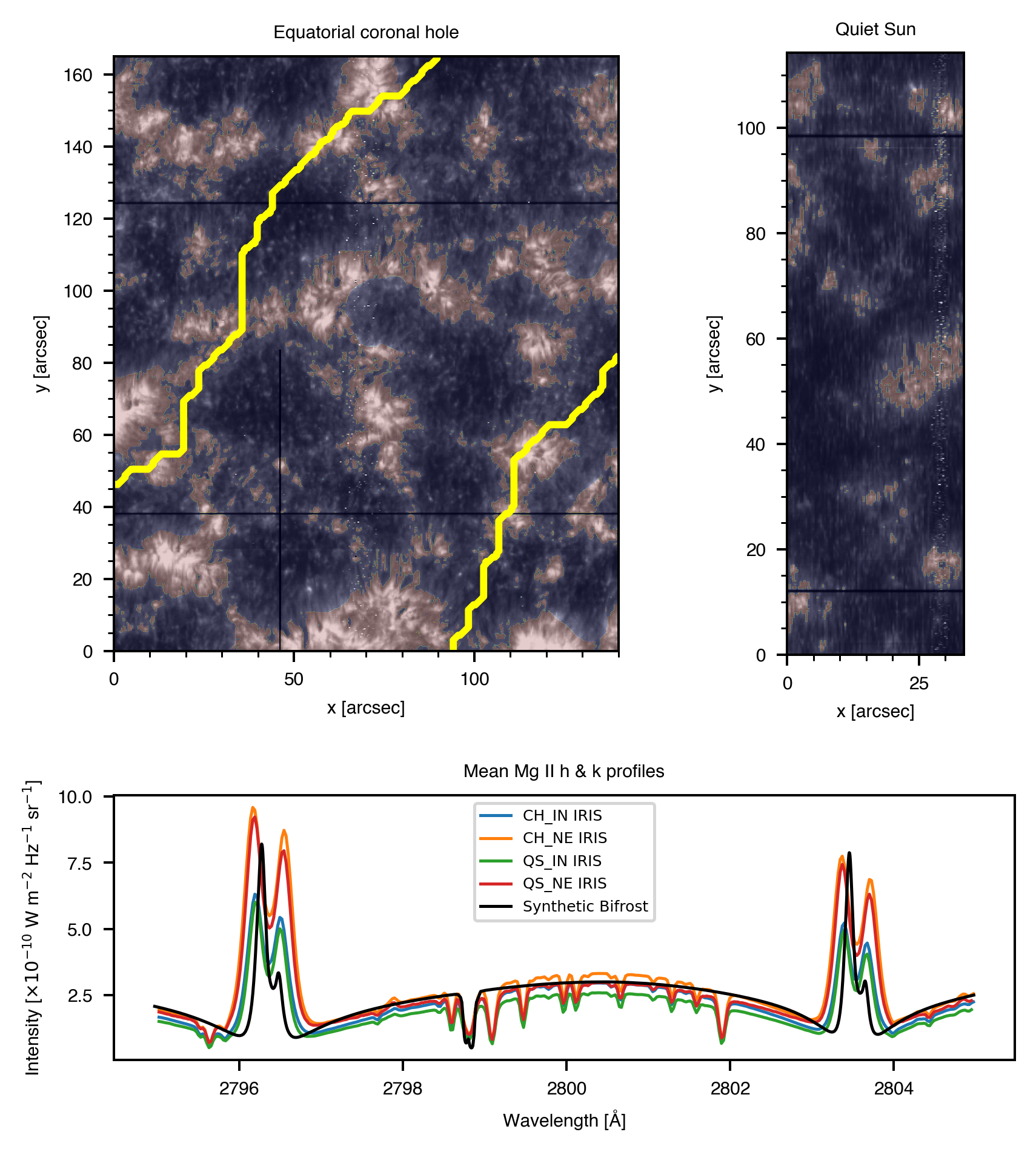}}
    \caption{\textit{Top panels:} Network (red shade) and internetwork (blue shade) regions as determined after applying the LCT technique on AIA 1600~\AA\/ filtergrams combined with intensity thresholds of an equatorial coronal hole (left) and the quiet Sun (right). For the coronal hole observation, the QS region has been removed (yellow countour). The background images are the corresponding IRIS rasters in the Mg II k line at 2796.6~\AA\/. Both rasters are scaled to the same value. \textit{Bottom panel:} Mean internetwork and network \mgii~h/k profiles in the selected equatorial coronal hole and the quiet Sun regions, and the mean synthetic Bifrost profile.}
    \label{fig:ch_qs_mask}
\end{figure*}

To compare synthetic with the observed \mgii~h/k spectral lines, we first separated IN from NE areas in the selected observed QS and CH regions, we assumed that IN coincides with the interior of supergranular cells. The NE areas enclose IN and is co-spatial with the boundaries of supergranules. Those boundaries are determined applying the local correlation tracking (LCT) technique \citep{November:1988oj} to AIA 1600~Å filtergrams, which are taken concurrently with the IRIS QS and CH sequences. The LCT algorithm calculated horizontal velocities employing a Gaussian tracking window of FWHM of 12\arcsec, which is large enough to suppress granular convective motions and preserve the large-scale flows. We identified the supergranular borders by tracking the evolution of passively advected tracers (corks). Since the solar magnetic fields may expand with height and our goal is to identify the chromospheric NE structures, we manually determined the cell interiors following the cork distributions and avoiding strong bright features co-spatial with the cell boundaries. Figure~\ref{fig:ch_qs_mask} shows NE regions (red shaded areas) around the blue shaded IN regions in the CH (left upper panel) and QS (right upper panel) environments.

While high S/N spectropolarimetric observations such as the ones required to validate the predicted synthetic observables are awaiting \dkist\ observations, intensity observations are already available for at least some of the spectral lines we have calculated. In the literature, the synthetic Stokes I profiles from numerical models are usually narrower than observed profiles, i.e., their width is smaller \citep[e.g.,][]{de-la-Cruz-Rodriguez:2012jb}. The reduced line width in the simulations is likely due to lower non-thermal or small-scale (unresolved) velocities in the atmosphere along the LOS, a lack of mass loading into the upper chromosphere, a lack of chromospheric heating, or a combination of these factors \citep{Carlsson:2015fk,Carlsson:2019ARA&A..57..189C}. Figure~\ref{fig:fts} compares the average synthetic profile from our simulation with the FTS atlas profile of \caii. Figure \ref{fig:ch_qs_mask}  compares the average synthetic profile with average \mgii\ profiles observed with \iris\ in the IN of CH and QS. The \iris\ instrumental spectral PSF has been taken into account when producing the synthetic profiles. The \caii\ profiles show quite a good match between observations and synthetic observables. The \mgii\ profiles do not agree that well, although their agreement in terms of intensity (also around the core) is good. In addition, our new IN simulation shows a better match in terms of width than any previous numerical simulation, even if they are still a bit too narrow, e.g., compare with the synthetic \mgii\ profiles in \citet{Carlsson:2019ARA&A..57..189C}.  We do note however that the synthetic profiles of \mgii\ have a substantial asymmetry of the k2 peaks suggesting a discrepancy in the thermodynamics of the upper chromosphere \citep{Leenaarts:2013ij,Pereira:2014eu}. This difference could in principle be caused by many issues. For example, our numerical domain is small ($6\times6$~Mm$^{2}$) and it lacks of any network or large field connectivity. Note that the profiles are taken for a single instance and for a small FOV. This means that the flow patterns may be dominated by a local transient/dynamic event, thus causing an asymmetrical profile. In addition, the field configuration is highly simplified and is missing the surrounding network field, which may produce a balance on the flux in the upper chromosphere. 

The comparison of observations with the simulation suggests that the chromospheric thermodynamics are relatively similar, suggesting that the local magnetic growth could be observed with DKIST capabilities. This will provide a new possible mechanism to feed the quiet chromosphere with a magnetic field and perhaps contribute to the heating. Below, we discuss in further detail which processes may be missing in the simulation that could lead to broader profiles. 

\section{Discussion and conclusions}\label{sec:dis}

\citet{Martinez-Sykora:2019dyn} presented a new possible mechanism for feeding the QS chromospheric magnetic field by converting the local chromospheric kinetic energy into magnetic energy in situ. They used a high-resolution self-consistent radiative MHD numerical model using the Bifrost code. The magnetic energy in the chromosphere is a crucial ingredient for heating and dynamics in the low solar atmosphere. Therefore, its origin and formation are critical for understanding these processes. Using this state-of-the-art model, we have synthesized photospheric and chromospheric Stokes profiles, which provide the necessary tools to compare with future observations. This comparison with observations may validate, constrain or test the simulations, or help to interpret the highly complex observations in the QS. The predicted signal seems to be large enough (especially the circular polarization) to be observed with \dkist. Those observations could discern if the local chromospheric magnetic field growth occurs in the Sun and, if so, in which types of regions.

Clear features in the synthetic observables highlight the differences in field topology between the photosphere and chromosphere. In the photosphere, the field is mostly localized near the downflow regions (dark in intensity), whereas in the chromosphere the magnetic field is larger in regions that show sharp transients in velocity and large Stokes I intensities at the line core. These magneto-thermodynamic observables both in the photosphere and low- and mid-chromosphere could be discerned with \dkist. The field seems to be more uniform through the chromosphere with some increase in the proximity of shocks. The dynamics associated with the shocks propagating through the chromosphere are responsible for the increase in magnetic field in the simulated QS \citep{Martinez-Sykora:2019dyn}.

Our analysis showed that, especially for chromospheric lines, it is critical to capture the magnetic observables while taking into account Doppler shift effects on each pixel in the field-of-view, rather than the typical approach of integrating the Stokes observables over a fixed wavelength range (i.e., ignoring variations in Doppler shift over the field-of-view). In addition, and in agreement with \citet{AfonsoDelgado2020_prep}, we find that the weak-field approximation or WFA provides good estimates of the magnetic field when the considered wavelength range is limited to one of the spectral features of the spectral lines, e.g., only the core of the lines. 

We also validated our simulation with existing \iris\ \mgii\ and FTS Atlas \caii\ observations. The FTS atlas data was obtained for QS including network field. For \iris, we selected IN regions in an equatorial CH and QS, both close to disk center. The \caii\ profile seems to agree quite well with the FTS atlas. For the \mgii\ profile the results are more mixed. The intensity agrees quite well, also at the core of the line. The synthetic \mgii\ line is narrower than the observations but the discrepancy is significantly reduced compared to previous observations. The asymmetry of the \mgii\ peaks however is too skewed towards the blue in comparison with the observations. This may indicate that some physical processes may be missing in the numerical model. For example, the radiative transfer calculations are based on the 1.5D approximation and are missing 3D radiative transfer effects. These have been shown to play a key role in the core of the \mgii\ line \citep{Judge2020ApJ...901...32J}. The narrower  \mgii\ core profile in the simulations could also come from a lack of turbulence or LOS velocity complexity, a lack of mass loading into the upper chromosphere, or a lack of heating. In our simulation we assumed that the plasma can be treated as a single fluid. Therefore, the ion-neutral interaction effects \citep[e.g., multi-fluid effects or ambipolar diffusion][]{Martinez-Sykora:2020ApJ...889...95M} are missing. Previous simulations have shown that these effects could lead to more heating and/or more mass loading into the chromosphere. It is also possible that other plasma physics processes could help reduce the discrepancy in line broadening or peak asymmetry; for example the Farley-Buneman and thermal instabilities, which could lead to micro-turbulence and heating \citep{Fletcher:2018ApJ...857..129F,Oppenheim2020ApJ...891L...9O,evans2022ApJprep}. Similarly, the numerical model assumes ionization in LTE, while it is known that non-equilibrium ionization plays a key role in the thermodynamics of the chromosphere. Including these physical processes in our numerical model is important to determine if the in-situ chromospheric magnetic energy growth can be dissipated into thermal energy via, for instance, ion-neutral interaction effects.

A more accurate atom model and more realistic conditions of the atmosphere and the radiation field (e.g., 3D effects and partial redistribution of the frequency) are critical components to properly reproduce the observed profiles from the synthesis of numerical simulations.  Finally, including the instrument's physical constraints (e.g., the instrumental profile of \visp\ at \dkist) will allow us to obtain closer synthetic profiles to the observed ones. In future work, we will include and test these parameters in our investigation.

\acknowledgements{\longacknowledgment} 

\bibliographystyle{aasjournal}
\bibliography{collectionbib.bib}

\end{document}